\documentclass[twocolumn,secnumarabic,amssymb, nobibnotes, aps, prl]{revtex4-2}

\usepackage[dvipdfmx]{graphicx}
\usepackage{color}

\newcommand{\red}[1]{{#1}}

\begin{document}

\title{Delayed onset and directionality of x-ray-induced atomic displacements \\ \red{observed on subatomic length scales}}

\author{Ichiro Inoue$^{1}$}%
\email{inoue@spring8.or.jp}
\author{Victor Tkachenko$^{2,3,4}$}
\email{victor.tkachenko@xfel.eu}
\author{Konrad J. Kapcia$^{5,4}$}
\author{Vladimir Lipp$^{2,4}$}
\author{Beata Ziaja$^{4,2}$}
\email{beata.ziaja-motyka@cfel.de}
\author{Yuichi Inubushi$^{1,6}$}
\author{Toru Hara$^{1}$}
\author{Makina Yabashi$^{1,6}$}
\author{Eiji Nishibori$^{7,8}$}
\email{nishibori.eiji.ga@u.tsukuba.ac.jp}

\affiliation{$^1$RIKEN SPring-8 Center, 1-1-1 Kouto, Sayo, Hyogo 679-5148, Japan.\\
$^2$Institute of Nuclear Physics, Polish Academy of Sciences, Radzikowskiego 152, 31-342 Krakow, Poland.\\
$^3$European XFEL GmbH, Holzkoppel 4, 22869 Schenefeld, Germany.\\
$^4$Center for Free-Electron Laser  Science CFEL, Deutsches Elektronen-Synchrotron DESY, Notkestr.  85, 22607 Hamburg, Germany.\\
$^5$Institute of Spintronics and Quantum Information, Faculty of Physics, Adam Mickiewicz University in Pozna$\acute{n}$, Uniwersytetu Pozna$\acute{n}$skiego 2, PL-61614 Pozna$\acute{n}$, Poland.\\
$^6$Japan Synchrotron Radiation Research Institute, Kouto 1-1-1, Sayo, Hyogo 679-5198, Japan.\\
$^7$Graduate School of Pure and Applied Sciences, University of Tsukuba, Tsukuba, Ibaraki 305-8571, Japan.\\
$^8$Faculty of Pure and Applied Sciences and Tsukuba Research Center for Energy Materials Science, University of Tsukuba, Tsukuba, Ibaraki 305-8571, Japan.}
%\date{}%
\begin{abstract}
Transient structural changes of Al$_2$O$_3$ \red{on subatomic length scales following} irradiation with an intense x-ray laser pulse (photon energy: 8.70 keV; pulse duration: 6 fs; fluence: \red{8$\times$10$^2$ J/cm$^{2}$}) have been investigated by using an x-ray pump x-ray probe technique. The measurement reveals that aluminum and oxygen atoms remain in their original positions by $\sim$20 fs after the intensity maximum of the pump pulse, followed by directional atomic displacements at the fixed unit cell parameters. By comparing the experimental results and theoretical simulations, we interpret that electron excitation and relaxation triggered by the pump pulse modifies the potential energy surface and drives the directional atomic displacements. Our results indicate that high-resolution x-ray structural analysis with the accuracy of 0.01 \AA\  is feasible even with intense x-ray pulses by making the pulse duration shorter than the timescale needed to complete electron excitation and relaxation processes, which usually take up to a few tens of femtoseconds.
\end{abstract}

\maketitle
Knowledge of the structure of matter at subatomic resolution is essential to understand and accurately predict material properties.  The interatomic distances measured with the accuracy of the order of 0.01 \AA, for example, give a clue to presume origins of chemical bonds (covalent, ionic, metallic, and van der Waals bonds) \cite{Cordero2008,Shannon1969,Shannon1976,Batsanov2001} as well as valence numbers of ions \cite{Shannon1969,Shannon1976} and spin states of atoms \cite{Cordero2008}.

X-ray scattering has been a primary tool for structural studies of various systems in physical, chemical, and biological sciences. The recent advent of x-ray free-electron lasers (XFELs) \cite{Saldin1999, McNeilNP2010}, which generate brilliant femtosecond x-ray pulses, is greatly enhancing the capabilities of x-rays as an atomic-resolution probe. One of the most important applications of XFEL radiation is the structure determination of nanocrystals \cite{Ilme2015}, which is challenging with conventional x-ray sources due to the radiation damage occurring during x-ray irradiation \cite{Owen2006,Howells2009,Garman2010,Holton2010}. When an XFEL pulse irradiates matter, photo-, Auger, and secondary electrons are emitted during or shortly after the pulse \cite{Ziaja2001, Ziaja2005}. Although the electron excitation ultimately triggers subsequent atomic disordering \cite{Medvedev2013, Medvedev2015, Medvedev2018}, it has been predicted \cite{Neutze2000} that ultrafast x-ray pulse can complete diffraction measurement before the onset of the atomic displacements \cite{Chapman2014}, and thereby overcome the radiation damage.

Even though the nominal duration of the x-ray pulses at most XFEL facilities (30 fs or longer \cite{Bostedt2016, Kang2017, Prat2020, Decking2020}) is longer than the predicted onset time of atomic disordering, structures of macromolecules were successfully determined in serial femtosecond crystallography (SFX) experiments \cite{Chapman2011, Boutet2012}. These successes are attributed to the self-terminating diffraction mechanism \cite{Barty2012}; x-ray-induced atomic disordering reduces the diffraction intensity, and the effective x-ray pulse duration becomes shorter than the original pulse duration. Theoretical simulations \cite{Jurek2009, Hau-Riege2015} and experiments \cite{Nass2015,Nass2020}, however, suggested that the atomic displacements are rather correlated than random, which may cause artifacts in the determined structures \cite{Lomb2011} because the diffraction intensity reflects the time-averaged structure over the effective pulse duration.

Detailed understanding of the mechanism for x-ray-induced atomic displacements is of great importance not only for confirming the validity of current SFX experiments, but for future realization of high-resolution structure analysis with intense XFEL pulses, such as visualizing charge-density distribution and chemical properties in matter at subatomic resolution ($\sim$0.01 \AA) \cite{Coppens1997}. A few pioneering groups attempted to follow x-ray-induced transient structural changes in various homoatomic materials (diamond \cite{Inoue2016, Inoue2021}, silicon \cite {Pardini2018,Hartley2021}, bismuth \cite{Makita2019}, and xenon clusters \cite{Ferguson2016}) and in protein crystals \cite{Nass2020} by using x-ray pump x-ray probe techniques. However, the initial disordering processes in these samples except for diamond \cite{Inoue2016, Inoue2021} have not been fully revealed due to the insufficient temporal resolution. Thus, there remain fundamental questions, such as  how x-ray-induced atomic displacements proceed with time and whether subatomic structure determination is feasible with intense XFEL pulses.

Here we describe intense x-ray-induced structural changes in a two-element compound, Al$_2$O$_3$, measured by an x-ray pump x-ray probe technique. The experimentally tested Al$_2$O$_3$ has a rhombohedral structure with space group $R\bar{3}c$. \red {As discussed below, the structure of the sample is very simple and can be described by only a few parameters, which enabled  structure determination with short measurement time and straightforward data analysis, and later, efficient theoretical simulations of the X-ray-induced transient structure changes.}
By employing a unique capability of SACLA \cite{IshikawaNP2012} that can generate XFEL pulses with duration of much below 10 fs \cite{Inubushi2017, InouePRAB2018, InoueJSR2019}, we succeeded to capture the onset of atomic displacements. Furthermore, the small inherent atomic disorder of Al$_2$O$_3$ \cite{Thompson1987} and the use of the probe pulse with a short wavelength allowed us to accurately determine the transient  atomic positions.

Figure 1 (a) shows a schematic illustration of the experiment at SACLA beamline 3 \cite{YabashiJSR2015}. The XFEL machine was operated in the split undulator mode \cite{Hara2013} to generate 8.70-keV pump and 11.99-keV probe pulses with the FWHM duration of 6 fs. The pump and probe pulses were focused to FWHM sizes of 2.2 $\mu$m (horizontal) × 2.7 $\mu$m (vertical) and 1.0 $\mu$m (horizontal) × 1.4 $\mu$m (vertical), respectively, by using an x-ray mirror system \cite{Yumoto2013, Tono2017}. A nanocrystal Al$_2$O$_3$ film (grain size of 80 nm, US research nanomaterials) with a thickness of 1 $\mu$m was placed at the focus and continuously translated spatially to expose an undamaged surface to each double pulse. The probe diffraction peaks in the scattering angle (2$\theta$) range of 16$^\circ$-90$^\circ$ were measured by using a multiport charge-coupled device detector \cite{KameshimaRSI2014}, while changing the delay time from 0.3 fs to 100 fs. Since the jitter of the delay time (several tens of attoseconds \cite{Hara2013}) was negligibly small, the duration of the XFEL pulse determined the time resolution of the measurement ($\sqrt{2}\times$6 fs =8.5 fs). Here, a 600-$\mu$m-thick aluminum foil in front of the detector absorbed the diffracted pump photons. The shot-by-shot pulse energy was characterized by an inline spectrometer \cite{Tamasaku2016}. More than 1500 pulses with the specific pulse energies of the pump (49.0 $\pm$ 5.4 $\mu$J) and the probe pulses (17.1 $\pm$ 8.6 $\mu$J) (the peak fluence of \red{8$\times10^2$ J/cm$^{2}$} and \red{1$\times10^3$ J/cm$^{2}$}, respectively) were extracted for each delay condition, and the corresponding diffraction images were used for the following analysis. For reference, the measurement with reduced pump fluence below structural damage threshold \cite{Medvedev2020} was also performed by inserting a 200-$\mu$m-thick silicon attenuator upstream of the focusing mirror (the peak fluence of the pump and probe pulses were \red{8$\times10^1$ J/cm$^{2}$}  and \red{4$\times10^2$ J/cm$^{2}$}, respectively).

Figure 1 (b) shows the integrated one-dimensional powder diffraction data (diffraction profile) for all delay times. The scattering angle of each reflection index was the same for all diffraction profiles. Moreover, no new diffraction peaks  appeared for the measured delay times. These observations indicate that the unit cell parameters (space group, symmetry operations, and lattice constants) remained the same as those before irradiation with the pump pulse. On the other hand, the diffraction peaks for higher diffraction angles (2$\theta >70^\circ$) almost disappeared  at longer delay times, indicating that the structural changes occurred on the femtosecond timescale.

%%%%%FIGURE 1%%%%%
\begin{figure}
\includegraphics[width=8.5cm]{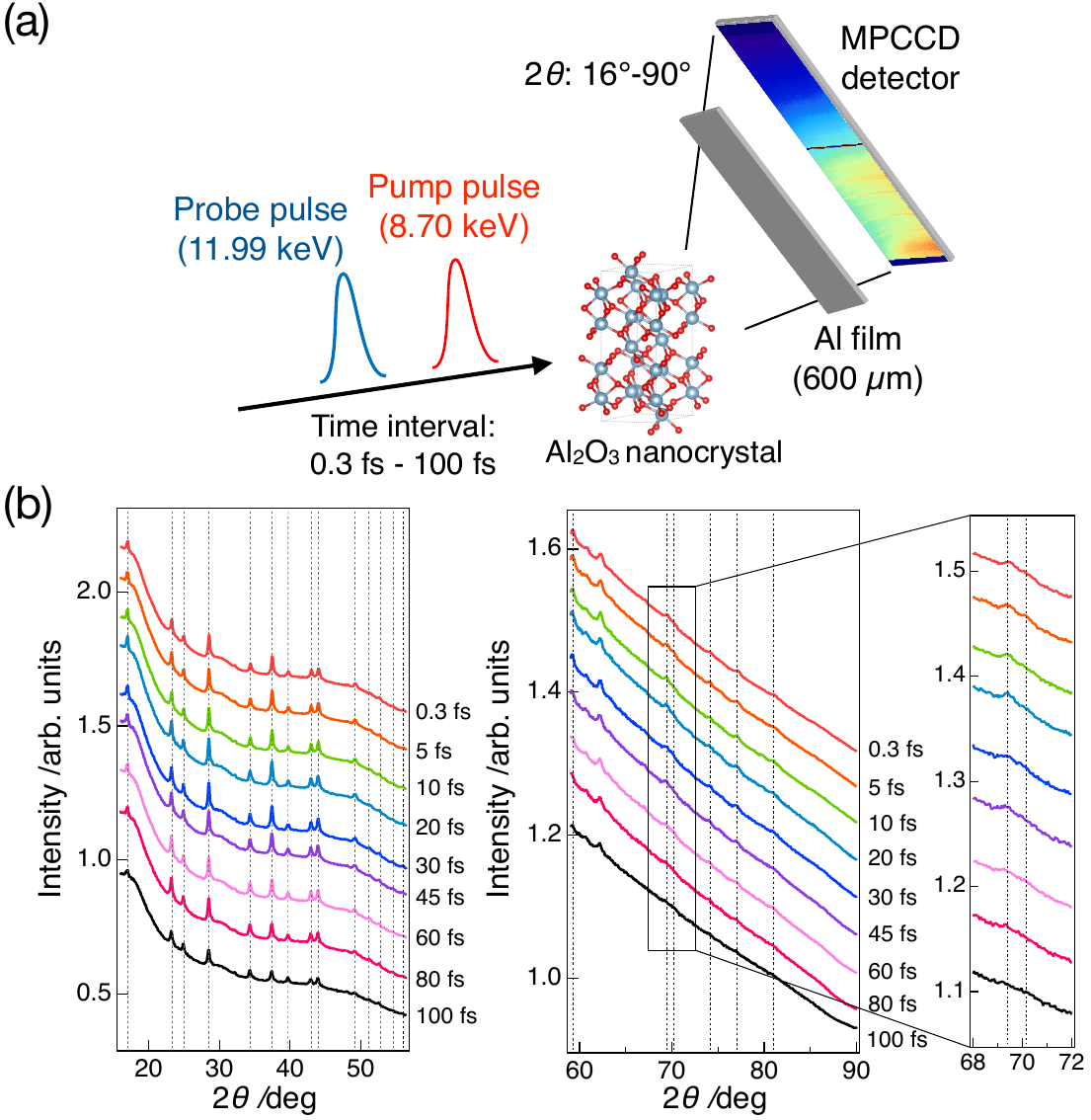}
\caption{(a) A schematic illustration of the experiment.
(b) Diffraction profiles for different delay times. For better visibility, each diffraction profile is linearly shifted along the vertical axis. Dotted lines show the positions of some of the diffraction peaks.}
\end{figure}
%%%%%FIGURE 1%%%%%

There are three possible reasons for this ultrafast decay of the diffraction intensity. The first is the change in atomic scattering factors due to progressing sample ionization. A theoretical calculation shows that the atomic scattering factors of the ionized atoms  at the measured scattering vectors are mostly determined by the number of occupied deep-shell levels \cite{Hau-Riege2007} and the valence electrons does not much contribute to the scattering intensity. As shown in the supplemental material \cite{SupplementalMaterial}, the number of deep-shell holes in both aluminum (Al) and oxygen (O) atoms was too small to explain the experimental observations. The second possible reason is the shift of atomic positions at the fixed unit cell parameters, i.e., the x-ray exposure might change the crystal-lattice sites. Such shift modifies relative phases of the scattering waves between different atoms and changes the diffraction intensity. Since the symmetry of the crystal structure was found to be preserved for the measured delay times, positions of all atoms within the unit cell can be defined by two parameters, fractional coordinate of Al atom along the $c$-axis ($z$) and that of O atom along the $a$-axis ($x$) (Fig. 2). The third possible reason of the observed signal decrease is the progressing atomic disorder, i.e.,  the positional fluctuation of atoms from the crystal-lattice sites might be increased by the x-ray exposure. Under the assumption that the atomic displacements from the crystal-lattice sites are isotropic and follow a Gaussian distribution, the atomic scattering factor decreases by a factor of  $\exp(-8\pi^2 \langle u^2 \rangle \sin^2 \theta/(3\lambda^2))$, where $\lambda$ is the x-ray wavelength and $\langle u^2\rangle$ is the mean squared displacement. The second and third reason will be considered in the following analysis.

%%%%%FIGURE 2%%%%%
\begin{figure}
\includegraphics[width=6cm]{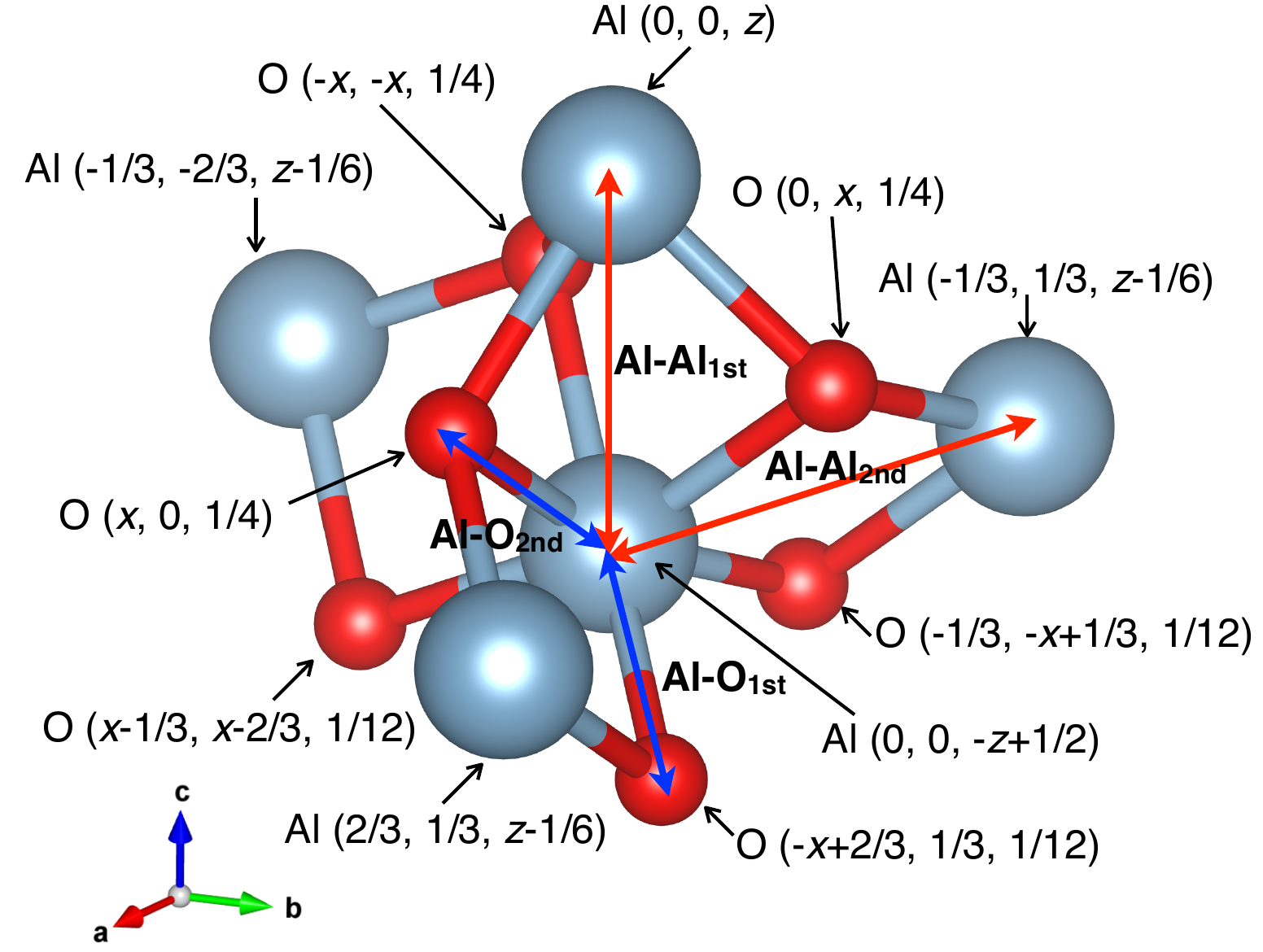}
\caption{Fractional coordinates of Al and O atoms around Al atom at (0, 0, -$z$+1/2). Al-Al$_{\rm{1st}}$ and Al-Al$_{\rm{2nd}}$ (Al-O$_{\rm{1st}}$ and Al-O$_{\rm{2nd}}$) represent the first and second neighboring distances between two Al atoms (Al and O atoms).}
\end{figure}
%%%%%FIGURE 2%%%%%

%%%%%FIGURE 3%%%%%
\begin{figure}
\includegraphics[width=7cm]{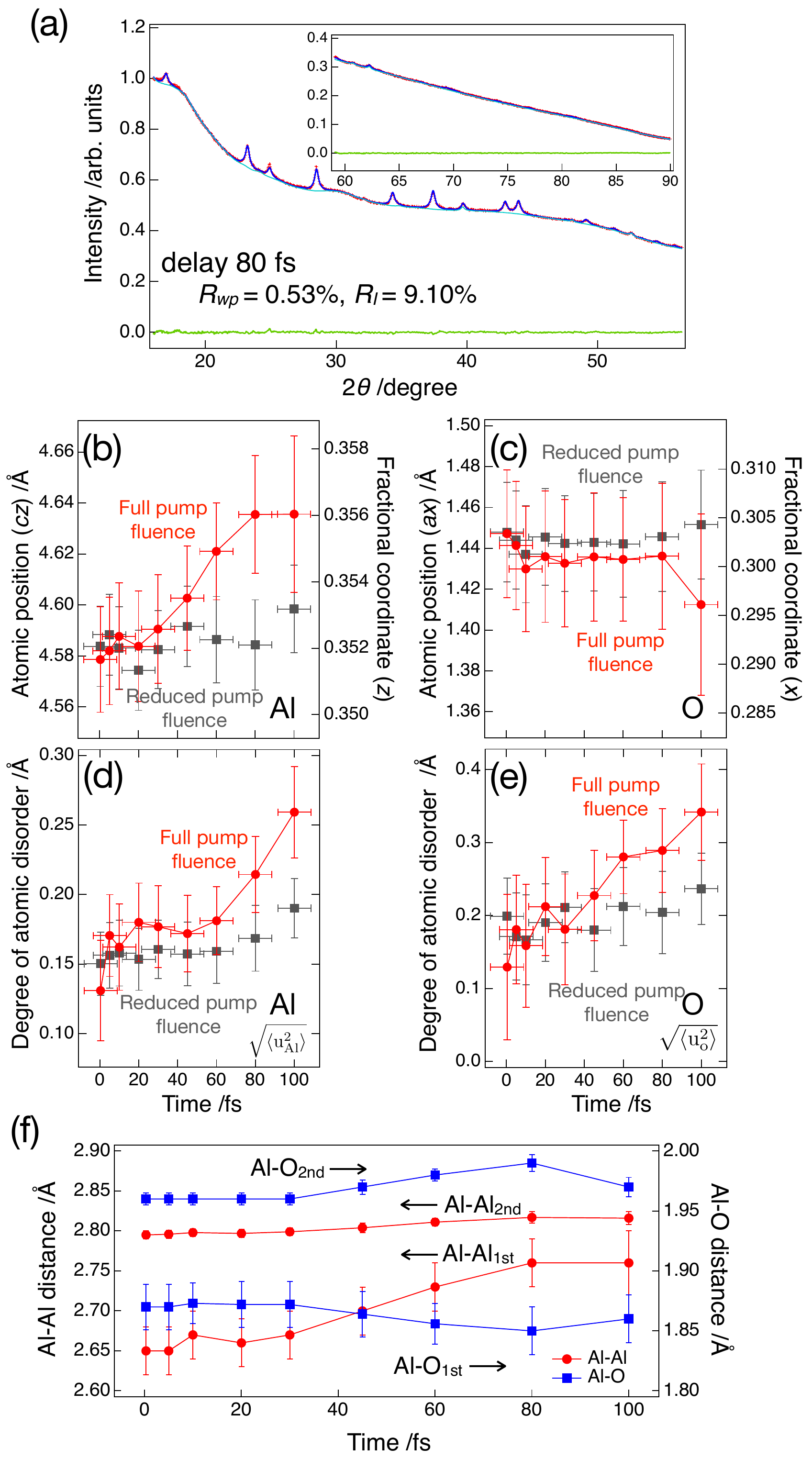}
\caption{(a) An example of Rietveld refinement. Red markers show the measured diffraction profiles. Blue, light blue, and green lines represent the fitted curve, the estimated background, and the deviation between the measured and fitted diffraction profiles, respectively.
(b-e) Atomic positions and root mean squared displacements of the atoms from the crystal-lattice sites after irradiation with the pump pulse with the fluence of \red{8$\times10^2$ J/cm$^2$} (red circles) and \red{8$\times10^1$ J/cm$^2$} (gray squares). Time zero corresponds to the intensity maximum of the pump pulse.
(f) Temporal changes of the interatomic distances (Al-Al$_{\rm{1st}}$, Al-Al$_{\rm{2nd}}$, Al-O$_{\rm{1st}}$, Al-O$_{\rm{2nd}}$) after irradiation with the pump pulse.
}
\end{figure}
%%%%%FIGURE 3%%%%%

In order to characterize the x-ray-induced structural change, Rietveld refinement was performed, assuming fixed lattice constants  ($a=b$ =4.77 \AA, $c$= 13.02 \AA, $\alpha=\beta=90^\circ$, $\gamma=120^\circ$) and isotropic positional fluctuation of atoms from the crystal-lattice sites.  Here we refined four structure parameters: (i), (ii) fractional coordinates of Al atom along the $c$-axis ($z$) and O atom along the $a$-axis ($x$), and (iii), (iv)  degree of atomic disorder for Al and O atoms ($\sqrt{\langle u^2_{\rm{Al}}\rangle}$  and $\sqrt{\langle u^2_{\rm{O}}\rangle}$), which is given by the root mean squared atomic displacement from the crystal-lattice sites. Rietveld refinement of the diffraction profiles yielded quite satisfactory fits (reliability factors $R_I$ and $R_{wp}$ \cite{Young1982} were $\sim$10\% and $\sim$0.5\%, respectively) (see Fig. 3(a) and the supplemental material \cite{SupplementalMaterial}) and the structure parameters as a function of time after the intensity maximum of the pump pulse were successfully determined (Figs. 3(b)-3(e)).  Notably, the positions of Al and O atoms within the unit cell (position of Al atom along the $c$-axis  $cz$ and that of O atom along the $a$-axis $ax$) were determined \red{with small errors of the orders of 0.01 \AA ($\sim$0.02 \AA\ and $\sim$0.04 \AA, respectively)} (Figs. 3(b) and 3(c)). For comparison, the structure parameters for the experiment with the reduced pump fluence below damage threshold are also shown in Figs. 3(b)-3(e).

It was found that the structure parameters remained almost the same and equaled to those obtained for the  reduced pump fluence until $\sim$20 fs after the intensity maximum of the pump pulse (Figs. 3(b)-3(e)), clearly indicating that the x-ray-induced atomic displacements were delayed in respect to the pump pulse. Interestingly, the atomic displacements were found to be highly directional; the positions of Al and O atom were clearly shifted along the $c$-axis and the $a$-$b$ plane, respectively, with time (Figs. 3 (b) and 3(c)) in accordance with the increase of the atomic disorder (Figs. 3(d) and 3(e)). The directional atomic displacements caused the changes in the interatomic distances (Fig.3 (f)). The results shown in Figs. 3(b)-3(f) confirm the need for the ultrashort X-ray pulses for accurate structure determination.

We can consider  two possible mechanisms for the directional atomic displacements. One possibility is that the pump excitation and subsequent processes increased the kinetic energy of atoms and made it possible for atoms to reach the positions at which the anharmonicity of the interatomic potential  was significant. Such anharmonicity may break symmetric atomic distribution around the equilibrium position and shift the atoms. However, given that the atomic disorder along a particular axis ($\sqrt{\langle u^2_{\rm{Al}}\rangle/3}$ and $\sqrt{\langle u^2_{\rm{O}}\rangle/3}$) was comparable to the shift of  atomic position, this mechanism seems  to be not  relevant. Another possible mechanism is that the electron excitation and relaxation induced by the pump pulse modified the  potential energy surface and drove atomic displacements, i.e., the directional atomic displacements were of a nonthermal origin. Previous simulations predict that the excited high-energy photo- and Auger electrons do not relax immediately after irradiation with the x-ray pulse. The respective electron cascading can take up to a few tens of fs \cite{Ziaja2001, Ziaja2005, Inoue2021, Son2011}. This timescale is consistent with the delayed onset of the atomic displacements  observed in the experiment.
    
        To confirm the validity of the delayed onset and the directionality of the atomic displacements, we simulated the time evolution of atomic positions and electronic structure within x-ray-excited Al$_2$O$_3$ by using \textit{XTANT+} (x-ray-induced thermal and nonthermal transitions plus) code \cite{Lipp2022}. The calculation was performed for a 240-atom-large supercell (96 Al atoms and 144 O atoms) irradiated with spatially uniform x-ray pulse with 6 fs duration (FWHM). From the results obtained at each time step, we determined structure parameters for Al$_2$O$_3$ ($x$, $z$, $\sqrt{\langle u^2_{\rm{Al}}\rangle}$, and $\sqrt{\langle u^2_{\rm{O}}\rangle}$), such that $\sqrt{\langle u^2_{\rm{Al}}\rangle}$ and $\sqrt{\langle u^2_{\rm{O}}\rangle}$ were minimized.
        
Figures 4(a)-4(d) show simulated structure parameters for three fluence values, corresponding to 100\%, 50\%, 10\%, and 0\% of the nominal peak fluence of the pump pulse in the present experiment (\red{8$\times$10$^2$ J/cm$^{2}$}). For the case of the 10\% nominal fluence, the structure parameters were almost the same as those without x-ray exposure (0\% fluence). For higher fluence values (100\% and 50\% fluence), the structure parameters start to deviate from those for 0\% fluence at $\sim$20 fs after the intensity maximum of the  x-ray pulse. The simulation predicts that the fractional coordinates of Al and O atoms change with time (Figs. 4(a) and 4(b)), which agree with the directional atomic displacements observed in the experiment. Fig. 4 (e) shows simulated electron and ion temperatures.  The ions remain “cold” before the x-ray-induced structural change, which supports our interpretation that the directional atomic displacements were of nonthermal origin and the onset time of the structural changes was governed by the timescale needed to complete electron excitation and relaxation. In fact, the simulated timescale of electron relaxation (20-30 fs after which the increase of the electron temperature saturates) is consistent with the onset time of the structural changes observed in the experiment. These simulation results and the experimental observations indicate that high-resolution structure determination at subatomic resolution \cite{Coppens1997}, which is a commonplace technique at synchrotron x-ray sources, is also feasible with intense XFEL pulses by making the pulse duration sufficiently shorter than the timescale of the electron excitation and relaxation.

%%%%%FIGURE 4%%%%%
\begin{figure}
\includegraphics[width=8cm]{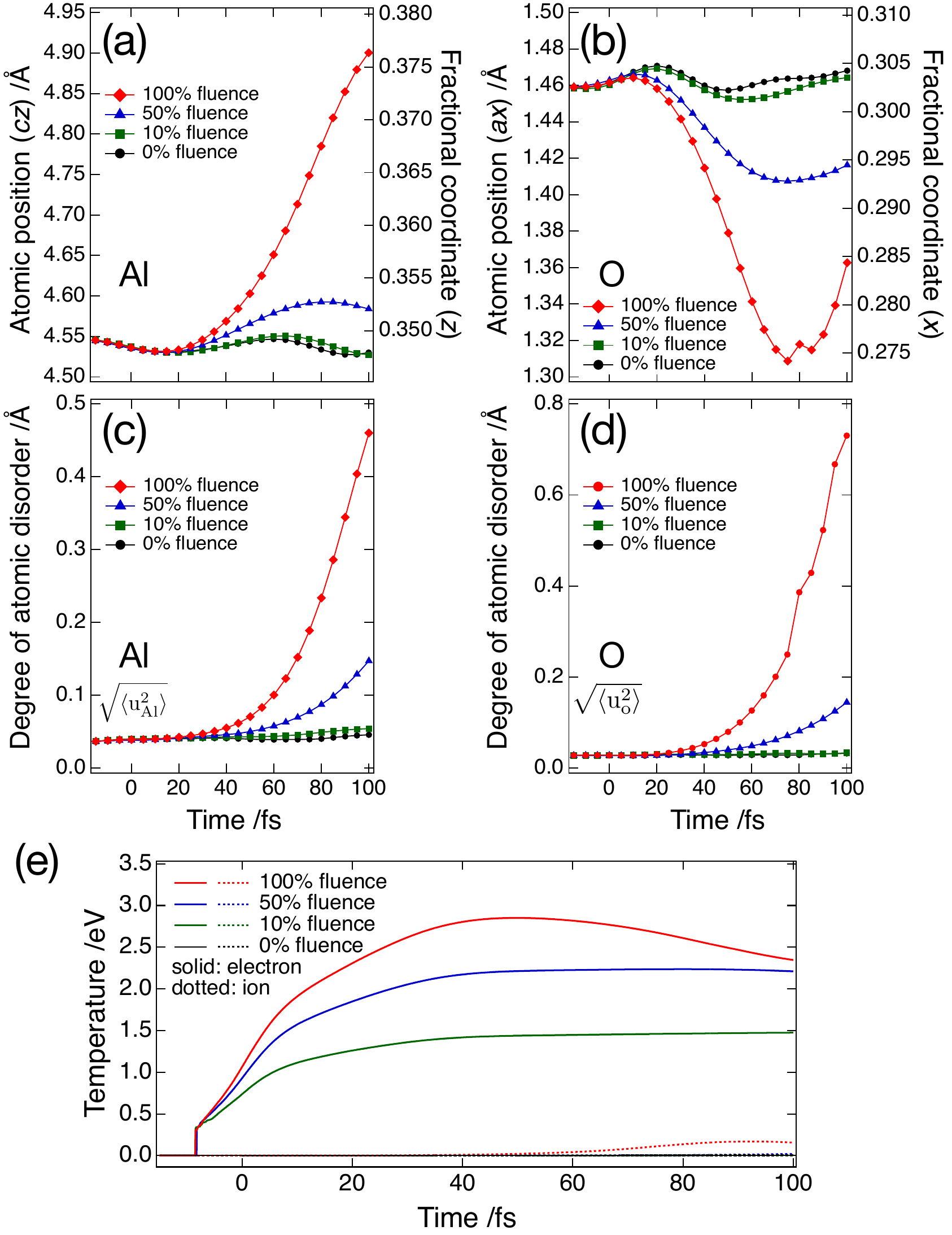}
\caption{(a)-(d) Theoretical predictions for structure parameters of Al$_2$O$_3$ after irradiation with an x-ray pulse. Time zero corresponds to the intensity maximum of the x-ray pulse.
(e) Simulated electron (solid lines) and ion (dotted lines) temperatures.
}
\end{figure}
%%%%%FIGURE 4%%%%%

Although the simulation for 100\% fluence reproduces the trend of the time-dependence of the structure parameters observed in the experiment, the predicted shift of the atomic positions and degree of the atomic disorder are larger than those evaluated by the experiment. This discrepancy can be explained by the non-uniformity of the pump fluence \cite{Victor2021}. Since the focal spots of the pump and probe pulses  had Gaussian shapes in the present experiment \cite{Yumoto2013}, the probe diffraction signals originated from various sample areas with different pump fluence. Figs. 4 (c) and 4(d) show that the predicted degree of atomic disorder increases with the pump fluence. Thus, we can consider that the scattering power of the crystals that experienced high pump fluence decayed faster and their contribution to the probe diffraction intensity decreased with time. Consequently, the “effective” pump fluence might be reduced with the increasing delay time in the present experiment, explaining the differences in the structure parameters for the experiment and the simulation.

\red{In summary, we investigated femtosecond structural changes of x-ray-excited Al$_2$O$_3$ on subatomic length scales.} It was found that the onset of the atomic displacements was delayed by $\sim$20 fs after the intensity maximum of the pump pulse, and that the atomic displacements were  directional rather than random. By comparing the experimental results and the theoretical simulation, we interpret that an  x-ray-induced change of the potential energy surface drove the directional atomic displacements. Since the nonthermal disordering would be ubiquitous for other covalent and ionic compounds, our results prove that high-resolution structure determination with the accuracy of the order of 0.01 \AA\ is feasible even with an intense x-ray by making the pulse duration shorter than the timescale of the electron excitation and collisional electron relaxation \cite{Ziaja2015}. Such advanced structure analysis provides new opportunities for x-ray science. For example, electron density mapping of charge-transfer processes in condensed matter upon photoexcitation could be an intriguing subject, accessible with such ultrashort pulses.

\acknowledgements{The experiments were performed with the approval of the Japan Synchrotron Radiation Research Institute (JASRI, Proposal No. 2018A8040). The work was supported by the Japan Society for the Promotion of Science (JSPS) KAKENHI Grants (19K20604, 19KK0132, 20H04656, and 21H05235). K. J. K. thanks the Polish National Agency for Academic Exchange for funding in the frame of the Bekker programme (PPN/BEK/2020/1/00184). Figures 1 (a) and 2 were drawn by using \textit{VESTA 3}\cite{Momma2011}}.

\bibstyle{natbib}
\bibliography{Ref}

\end{document}

\title{Supplemental Material: Delayed onset and directionality of x-ray-induced atomic displacements \red{observed on subatomic length scales}}
\author{Ichiro Inoue$^{1}$}%
\email{inoue@spring8.or.jp}
\author{Victor Tkachenko$^{2,3,4}$}
\email{victor.tkachenko@xfel.eu}
\author{Konrad J. Kapcia$^{5,4}$}
\author{Vladimir Lipp$^{2,4}$}
\author{Beata Ziaja$^{4,2}$}
\email{beata.ziaja-motyka@cfel.de}
\author{Yuichi Inubushi$^{1,6}$}
\author{Toru Hara$^{1}$}
\author{Makina Yabashi$^{1,6}$}
\author{Eiji Nishibori$^{7,8}$}
\email{nishibori.eiji.ga@u.tsukuba.ac.jp}

\affiliation{$^1$RIKEN SPring-8 Center, 1-1-1 Kouto, Sayo, Hyogo 679-5148, Japan.\\
$^2$Institute of Nuclear Physics, Polish Academy of Sciences, Radzikowskiego 152, 31-342 Krakow, Poland.\\
$^3$European XFEL GmbH, Holzkoppel 4, 22869 Schenefeld, Germany.\\
$^4$Center for Free-Electron Laser  Science CFEL, Deutsches Elektronen-Synchrotron DESY, Notkestr.  85, 22607 Hamburg, Germany.\\
$^5$Institute of Spintronics and Quantum Information, Faculty of Physics, Adam Mickiewicz University in Pozna$\acute{n}$, Uniwersytetu Pozna$\acute{n}$skiego 2, PL-61614 Pozna$\acute{n}$, Poland.\\
$^6$Japan Synchrotron Radiation Research Institute, Kouto 1-1-1, Sayo, Hyogo 679-5198, Japan.\\
$^7$Graduate School of Pure and Applied Sciences, University of Tsukuba, Tsukuba, Ibaraki 305-8571, Japan.\\
$^8$Faculty of Pure and Applied Sciences and Tsukuba Research Center for Energy Materials Science, University of Tsukuba, Tsukuba, Ibaraki 305-8571, Japan.}
\maketitle
\subsection{1. Results of Rietveld refinement}

 To characterize the x-ray-induced structure change of Al$_2$O$_3$, Rietveld refinement was performed under the assumption of the fixed lattice constants
 ($a=b$ =4.77 \AA, $c$= 13.02 \AA, $\alpha=\beta=90^\circ$, $\gamma=120^\circ$)  and isotropic atomic disordering.
   Here we refined four structure parameters: (i), (ii) fractional coordinates of Al atom along the $c$-axis ($z$) and O atom along the $a$-axis ($x$), and (iii), (iv)  degree of atomic disorder for Al and O atoms ($\sqrt{\langle u^2_{\rm{Al}}\rangle}$  and $\sqrt{\langle u^2_{\rm{O}}\rangle}$ ), which is given by the root mean squared atomic displacement from the crystal-lattice sites.
 Figure S1 shows the results of Rietveld refinement. In the figure, the red markers show the measured diffraction profiles. The blue, light blue, and green lines represent the fitted curve, the estimated background, and the deviation between the measured and fitted diffraction profiles, respectively.
$R_I$ and $R_{wp}$  in each plot are intensity and weighted reliability factors \cite{Young1982}.

\begin{figure}[h]
\includegraphics[width=13.6cm]{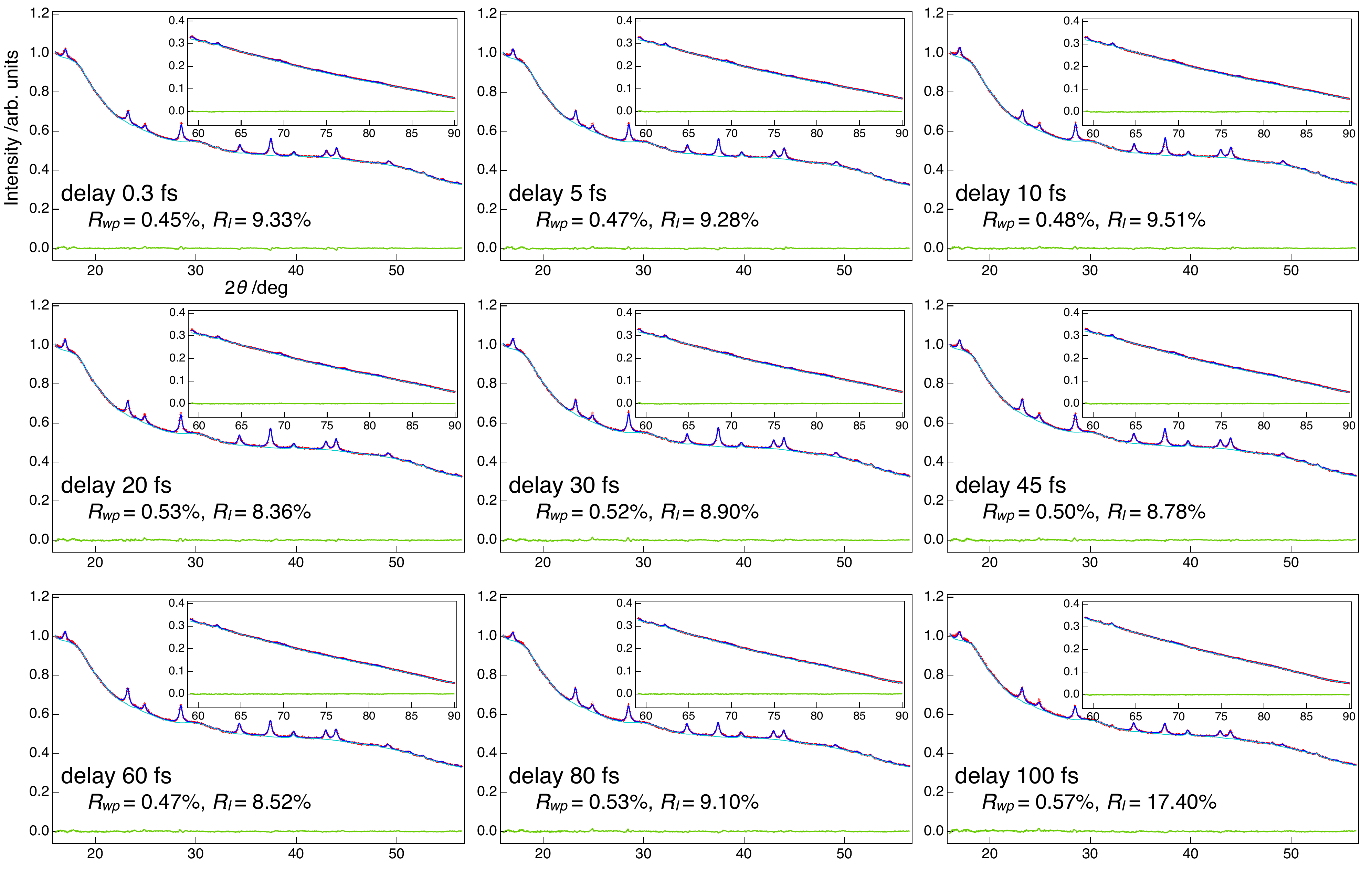}
\caption{Fitting results of Rietveld refinements for x-ray-excited Al$_2$O$_3$.}
\end{figure}

\red{Figure S2 shows the absolute values of the measured structure factors ($|F|$) as a function of diffraction angle $2\theta$.  The structure factors for the shorter delay times (0.3, 5, 10 fs) were almost the same, indicating that the structure of Al$_2$O$_3$ did not change on this time scale. In contrast, the structure factors for the longer delay times were  different from those for the shorter delay times. Particularly, $|F|$  for higher diffraction angles ($2\theta >70^\circ$) significantly decreased with the delay time due to the ultrafast structural changes.}

\begin{figure}[h]
\includegraphics[width=11cm]{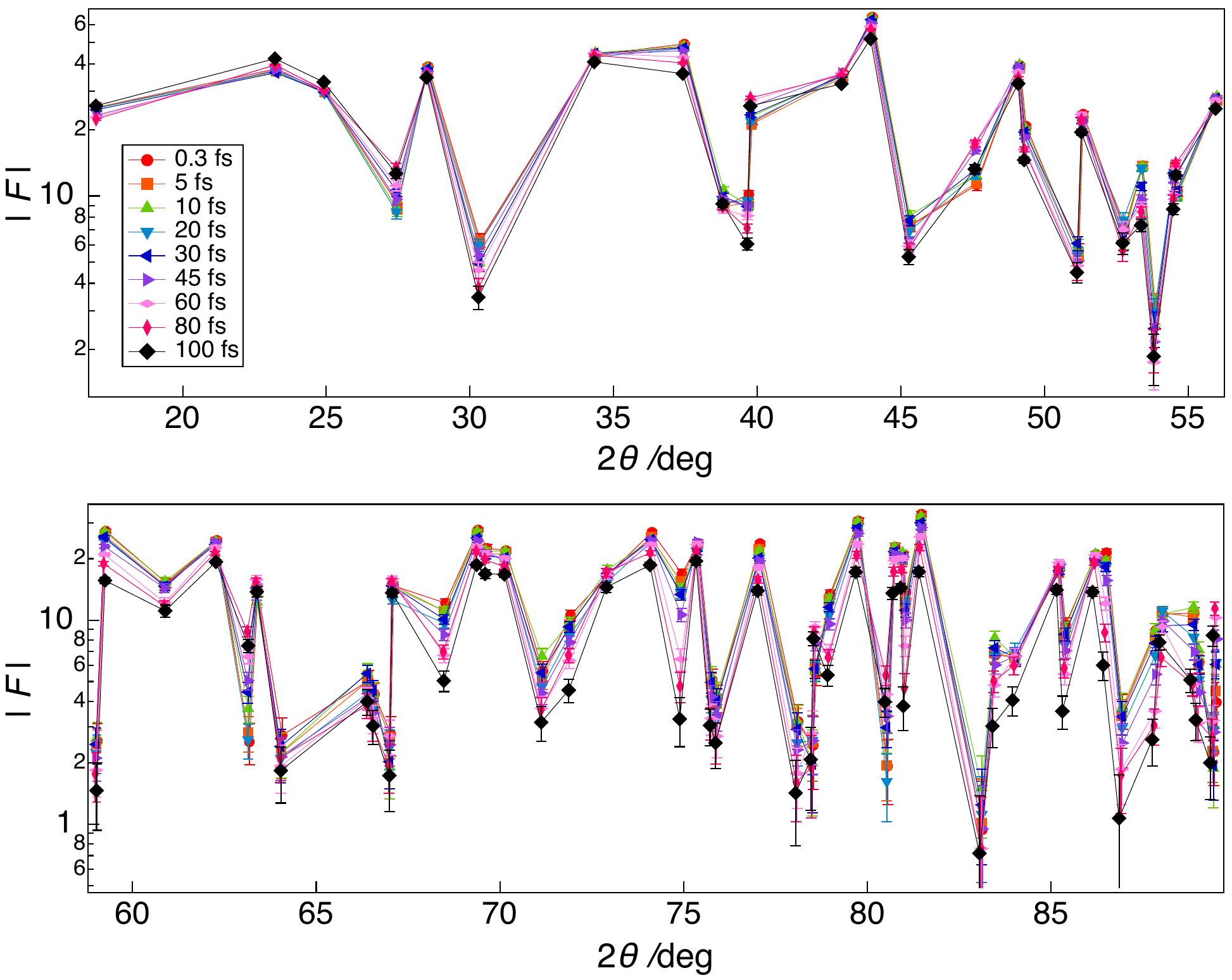}
\caption{\red{Absolute values of the measured structure factors as a function of diffraction angle.}}
\end{figure}

\clearpage

\subsection{2. Effect of deep-shell holes on diffraction intensity}
To evaluate the effect of the electron excitation on the diffraction intensity, we estimated number of deep-shell holes in aluminum (Al) and oxygen (O) atoms by  using \textit{XTANT+} code.
The calculation was  performed for a 240-atom-large supercell (96 Al atoms and 144 O atoms) irradiated with a spatially uniform x-ray pulse with duration of 6 fs (FWHM).
Figure S3 shows the calculated number of deep-shell holes per atom for three fluence values, corresponding to 100\%, 50\%, and 10\% of the nominal peak fluence of the pump pulse in the present experiment (\red{8$\times$10$^2$ J/cm$^{2}$}).
Simulation shows that the number of $K$-shell holes of O atoms is much less than 0.001 for the simulated fluence range,  while total number of deep-shell holes in Al atoms is estimated to be less than 0.03 per atom within the simulated time scale.
Given that the atomic scattering factors of the ionized atoms  at the measured scattering vectors are mostly determined by the number of occupied deep-shell levels and that the valence electrons does not much contribute to the scattering intensity \cite{Hau-Riege2007}, the estimated numbers of deep-shell holes correspond to the change in the diffraction intensity of much less than 1\%, which is too small to explain the transient changes in the diffraction intensity at the measured delay range.

\begin{figure}[h]
\includegraphics[width=13cm]{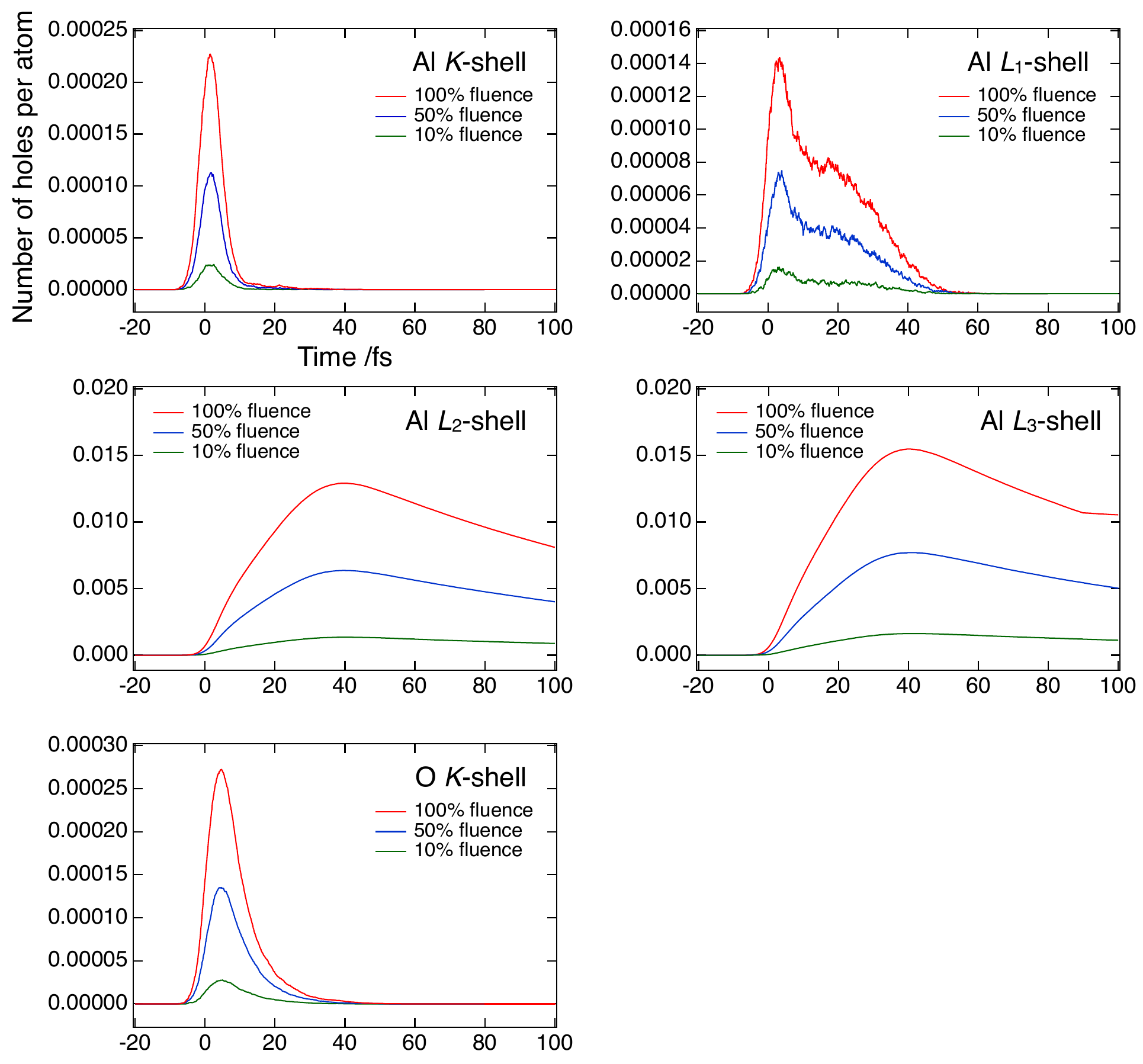}
\caption{Theoretical predictions for the number of holes per atom at various shells after irradiation with an x-ray pulse. Time zero corresponds to the intensity maximum of the x-ray pump pulse.}
\end{figure}

\bibstyle{natbib}

\bibliography{Ref}